\documentclass[11pt,twoside]{article}

\usepackage{asp2004}
\usepackage{epsf}
\usepackage{psfig}

\begin{document}

\title{The Ionized Component of Intermediate-Velocity Cloud Complex L}
\author{L. M. Haffner}
\affil{Department of Astronomy, University of Wisconsin--Madison}

\begin{abstract}
We present the first full map of the H\ensuremath{\alpha}\ emission from the inter\-mediate-velocity cloud Complex L. Kinematically resolved emission components from the ionized gas reveal a close spatial and velocity correspondence to that of the neutral gas as mapped by 21 cm line observations. Using a simple Gaussian component fitting technique, we compare the details of the emission from the two phases of Complex L. The mean velocity of the H\ensuremath{\alpha}\ and 21 cm emission are strongly correlated, but we find that the some of the ionized gas is systematically offset ($\sim -4$ km~s\ensuremath{^{-1}}) relative to the neutral gas. Line widths are qualitatively comparable, but both emission components are too low signal-to-noise to study any differences between the phases. Finally, as with our previous study toward Complex K, we find \ensuremath{I_{\mathrm{H}\alpha}}\ is not strongly correlated to \ensuremath{N_{\mathrm{H\;I}}}\ along the same lines of sight through Complex L. 
\end{abstract}

\section{Introduction}
Intermediate- and high-velocity clouds (IVCs and HVCs) are routinely being studied in optical emission lines thanks to advances in detector and instrument design (\citealp{Putman+03, Tufte+02}; \citealp*{WVW02,HRT01,TRH98}). In particular, the efficiency of the Wisconsin H-Alpha Mapper (WHAM) allows a multi-wavelength study of the ionized component over the entire extent of these predominantly neutral structures. The relationship between the H\ensuremath{\alpha}\ and 21 cm emitting gas should provide important clues into the source of ionization and conditions within the Galactic halo. Lyman continuum radiation is believed to escape the disk with sufficient flux to maintain the warm ionized medium (WIM) of the Milky Way and thus may also ionize the downward facing surfaces of neutral halo material (\citealp{WVW02,B-HM99}; \citealp*{DSF00,MC93}). However, high ionization species are also detected toward these complexes, suggesting that alternative forms of ionization and sources of Lyman continuum radiation may be important. Conductive interfaces or turbulent mixing layers arising from the interaction of the neutral gas with a very low density halo component have often been cited as mechanisms able to complete the picture of these anomalous structures \citep{Fox+04}.

We have embarked on an extended multi-wavelength study of the large-scale complexes with WHAM to explore the physical conditions in the ionized gas of IVCs and HVCs and to study the relationship of this component to the neutral gas. Here we present the first map of the ionized component of Complex L, first designated by \citet{WvW91} to be 21 cm emission present between $-190$ km~s\ensuremath{^{-1}}\ $< \ensuremath{v_{\mathrm{LSR}}} < -85$ km~s\ensuremath{^{-1}}\ in the approximate range $341\deg < \ell < 348\deg$ and $31\deg < b < 41\deg$. In this work, we examine a slightly larger region of the sky that shows intermediate-velocity emission in both 21 cm and H\ensuremath{\alpha}\ at slightly higher latitudes and longitudes that may be physically related to the gas previously defined to be Complex L.

There have been a few previous detections of H\ensuremath{\alpha}\ emission from Complex L. \citet*{WVW01} and \citet{Putman+03} find emission ($\ensuremath{I_{\mathrm{H}\alpha}} \sim$ 0.2 -- 1 R) toward several directions near higher column-density clumps of the complex. These emitting regions are also reported to have very high [\ion{N}{II}]/H\ensuremath{\alpha}\ ratios (1 -- 2.7). Here, we present the first map of the entire ionized component of Complex L.

\section{Observations}

All 21 cm data presented in this work was extracted from the Lieden/Dwingeloo survey \citep[LDS;][]{HIAtlas} and has a resolution of 0\fdg5 on the sky and a spectral resolution of 1 km~s\ensuremath{^{-1}}. A portion of the lower-velocity H\ensuremath{\alpha}\ emission from Complex L is present in the WHAM Northern Sky Survey (WHAM-NSS). See \citet{WHAMNSS} for details of WHAM and the survey.

Additional H\ensuremath{\alpha}\ data was obtained in 2004 April toward this region of the sky at the same pointing directions as the WHAM-NSS but with the velocity window shifted to capture $-250$ km~s\ensuremath{^{-1}} $< \ensuremath{v_{\mathrm{LSR}}} < -50$ km~s\ensuremath{^{-1}}\ and with twice the exposure time (60 sec per pointing). These data were processed, calibrated, and cleaned of atmospheric emission using a procedure very similar to that used for the NSS \citep[see][]{WHAMNSS}. Additionally, we have applied an estimated correction for the attenuation of H\ensuremath{\alpha}\ emission within each beam using the integrated \ensuremath{N_{\mathrm{H\;I}}}\ between $-50$ km~s\ensuremath{^{-1}} $< \ensuremath{v_{\mathrm{LSR}}} < +100$ km~s\ensuremath{^{-1}}, the relationship between \ensuremath{N_{\mathrm{H\;I}}}\ and extinction found by \citet*{BSD78}, and the analytic extinction law provided by \citet*{CCM89}. The number of assumptions in this correction make the resulting intrinsic H\ensuremath{\alpha}\ intensities only approximate. However, we hope to remove at least some of the apparent variation in the emission that may be due to foreground extinction. The average correction for the pointings in this region is $\sim1.4$.

For the results presented here, the two H\ensuremath{\alpha}\ spectra (with different velocity ranges) for each pointing were not combined into a single spectrum. Instead, except for one of the velocity-integrated channel maps presented below, the analysis focuses on using the newly taken data where the intermediate- and high-velocity emission is better placed within each spectrum. 

\begin{figure}[tbh]
\begin{center}
\plottwo{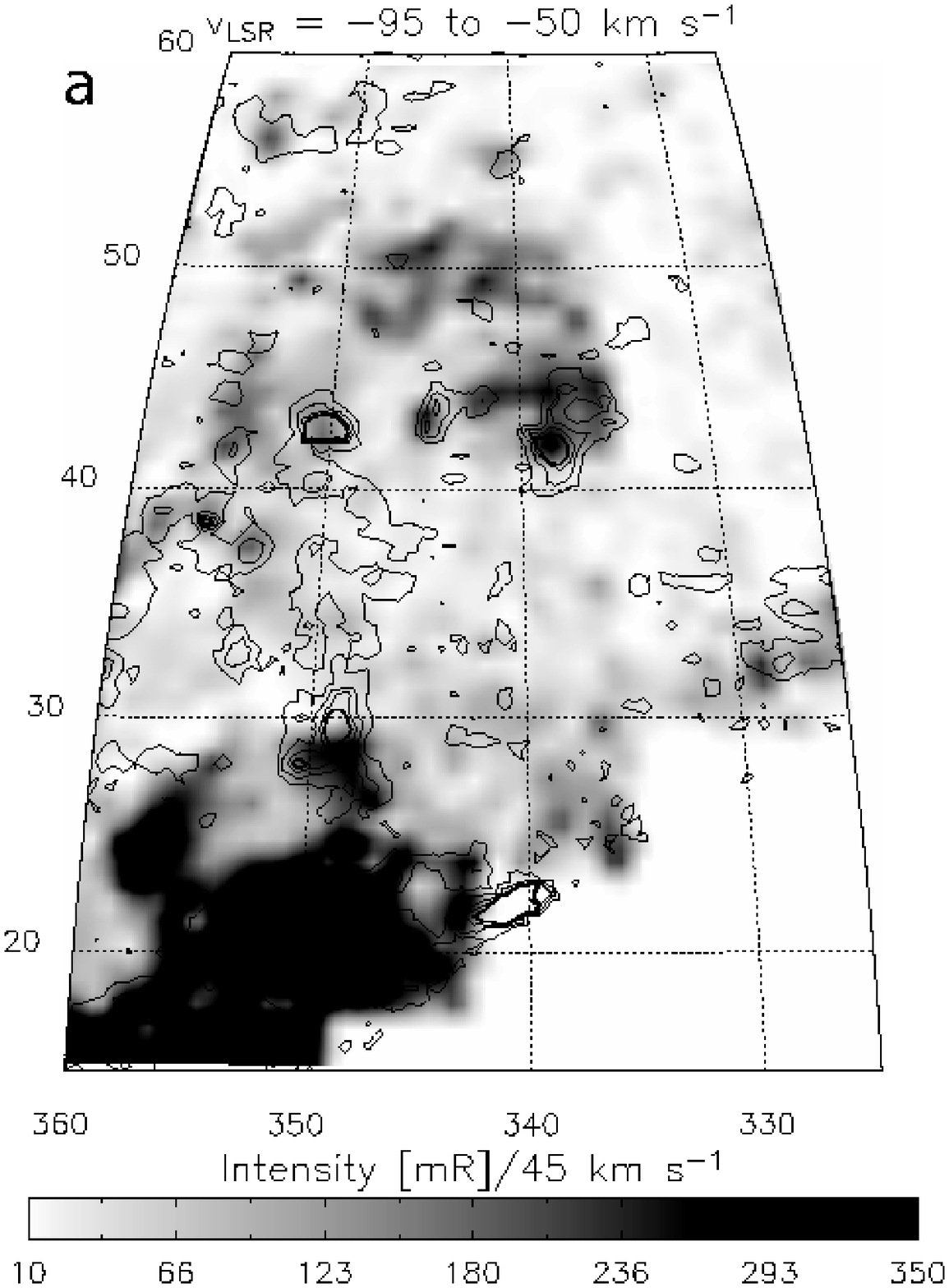}{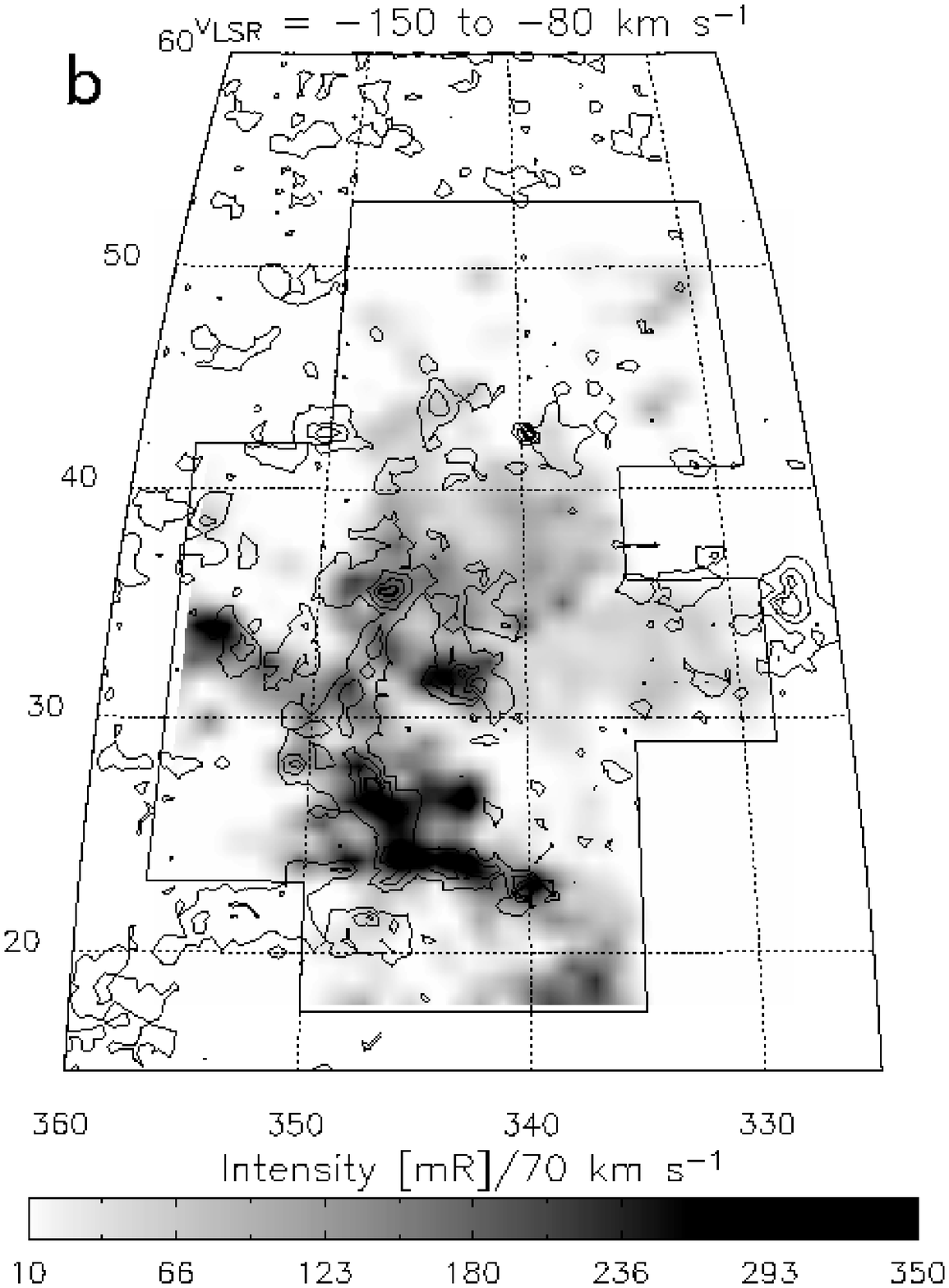}
\caption{H\ensuremath{\alpha}\ and 21 cm emission from Complex L. The grayscale image in each figure traces \ensuremath{I_{\mathrm{H}\alpha}}\ with levels denoted in the intensity bar at the bottom of each plot. Contours display \ensuremath{N_{\mathrm{H\;I}}}\ at levels of 6, 12, 18 and $24 \times 10^{18}$ cm$^{-2}$ from the LDS. \emph{(a)} shows the integrated emission from $-95$ km~s\ensuremath{^{-1}}\ $< \ensuremath{v_{\mathrm{LSR}}} < -50$ km~s\ensuremath{^{-1}}. The H\ensuremath{\alpha}\ data comes from the WHAM-NSS. Both the LDS and the WHAM-NSS end near $\delta = -30\deg$, which crosses the lower-right portion of the figure. \emph{(b)} shows the integrated emission from $-150$ km~s\ensuremath{^{-1}}\ $< \ensuremath{v_{\mathrm{LSR}}} < -80$ km~s\ensuremath{^{-1}}. The black line denotes the limit of the new, higher sensitivity and more negative velocity observations obtained with WHAM for this study.}
\label{fig1}
\end{center}
\end{figure}
 
\section{Results}

Figure~\ref{fig1} shows two velocity-integrated channel maps of H\ensuremath{\alpha}\ emission toward the region of the sky containing Complex L. Contours outline 21 cm emission from the neutral gas in the region. The bulk of the emission in both lines arises from distinct velocity components centered within the integrated ranges except for the brighter emission at $b < +25\deg$ in Figure~\ref{fig1}a, which is contamination from the wing of very bright, low-velocity emission in the $\delta$ Sco \ion{H}{II} region ($\ensuremath{I_{\mathrm{H}\alpha}} > 15$ R; $\ensuremath{v_{\mathrm{LSR}}} \sim -15$ to $-5$ km~s\ensuremath{^{-1}}). In both maps, the H\ensuremath{\alpha}\ and 21 cm emission are present over the same general regions and have similar spatial extents. But as with Complex K \citep{HRT01}, the intensities of each line do not trace each other in detail. 

In comparison with the earlier pointed observations made toward Complex L by \citet{Putman+03} and \citet{WVW01}, our measurements toward similar directions are roughly consistent but systematically lower. In particular, the WHAM observations made here with one-degree resolution do not show emission levels as high as those found by \citet{WVW01}. However, both of these studies have effective beam sizes smaller than WHAM by factors of 6--10. If significant variation of \ensuremath{I_{\mathrm{H}\alpha}}\ occurs on smaller scales, these measurements are not necessarily inconsistent.

\begin{figure}[tbh]
\begin{center}
\plottwo{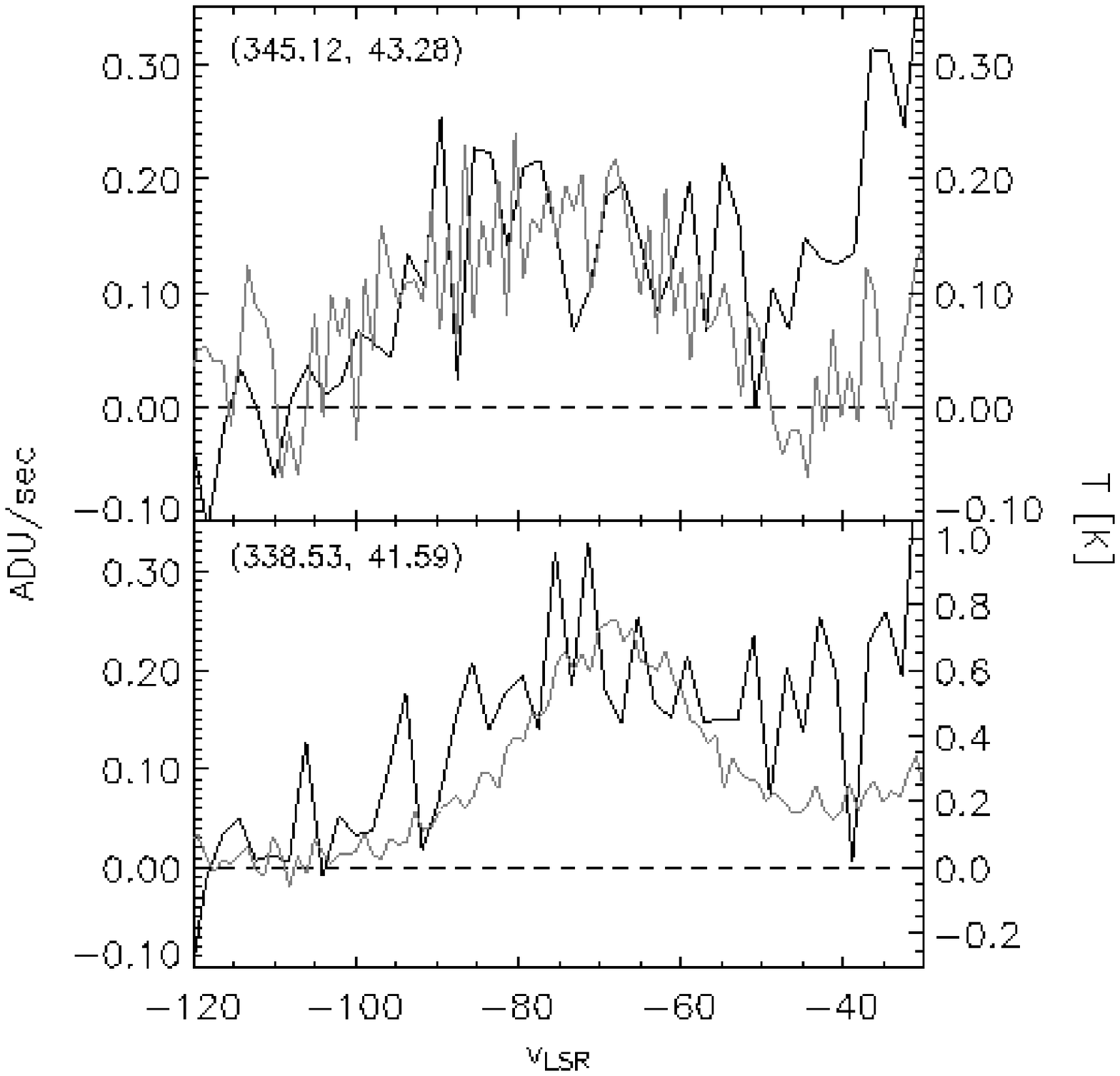}{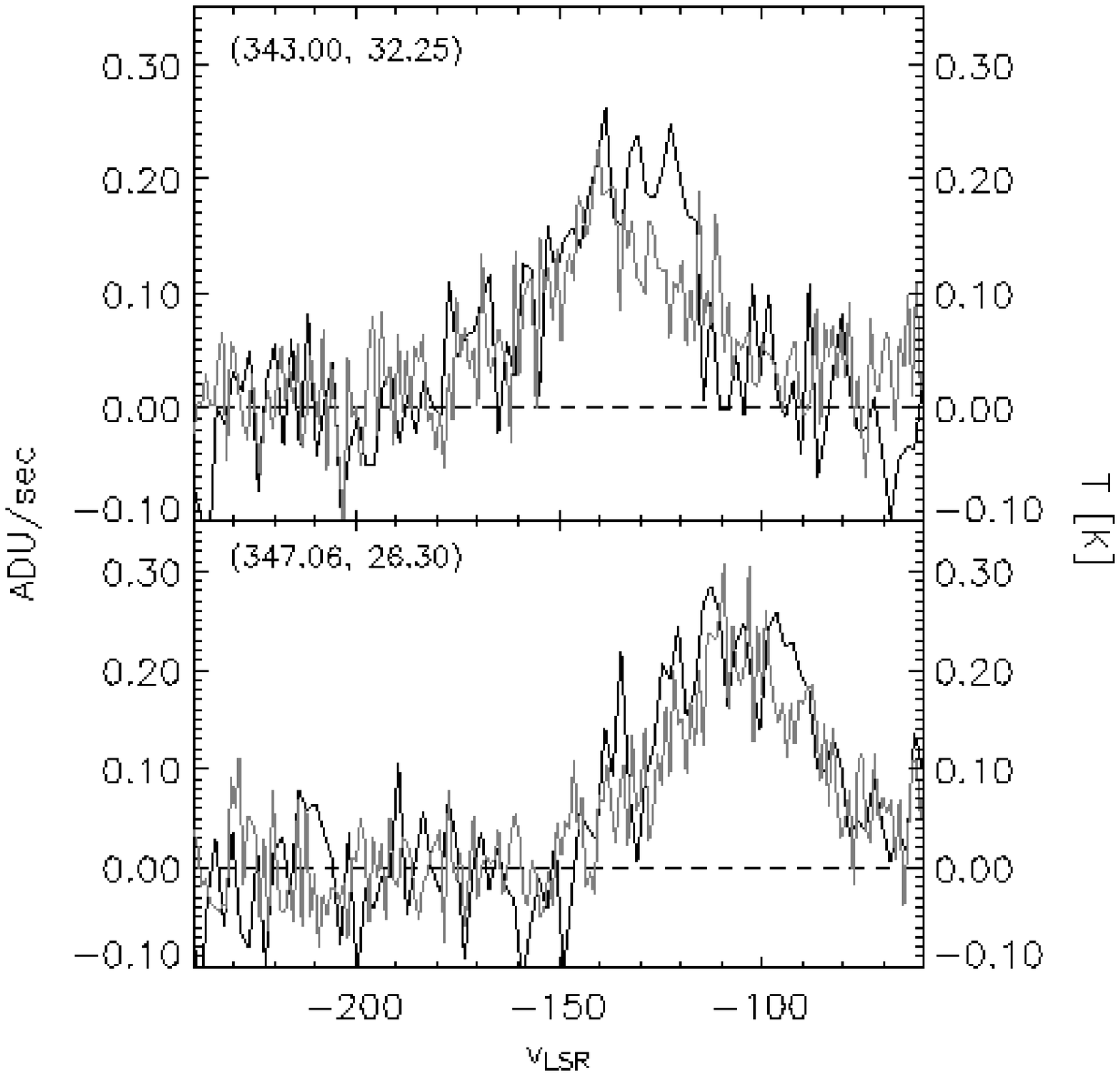}
\caption{Sample spectra from Complex L. Four sample beams from the data used to create Figure~\ref{fig1} are depicted here. The black line and left-hand y-axis plot the H\ensuremath{\alpha}\ emission profile for each one-degree beam. The gray line and right-hand y-axis trace the 21 cm emission for an average spectrum computed from LDS beams that fall within the one-degree WHAM pointing denoted by the coordinates,  ($\ell, b$), specified in the top-left of each panel.}
\label{fig2}
\end{center}
\end{figure}
 
Figure~\ref{fig2} shows sample spectra from the ionized and neutral components of Complex L. The H\ensuremath{\alpha}\ emission profile is displayed along with an average 21 cm spectrum constructed from LDS beams whose centers fall within the one-degree diameter WHAM beam. Both the location of the components and their velocity extents are quite similar. Although the relative intensities between the components change substantially over the area of Complex L, the similar location and shape of the emission profiles are a constant feature.

To explore the nature of this connection more quantitatively, we developed a straightforward single-Gaussian, auto-fitting program. Since the majority of the component emission from Complex L is well-separated from local emission features, a reasonable number of the profiles can be easily fit with such a routine. For our first pass at such an analysis, we followed the following prescription:

\begin{itemize}
\item Only fit if the integrated intensity is $\ensuremath{I_{\mathrm{H}\alpha}} > 50$ mR or $\ensuremath{N_{\mathrm{H\;I}}} > 2 \times 10^{18}$
\item Fit over the range $-200$ km~s\ensuremath{^{-1}} $< \ensuremath{v_{\mathrm{LSR}}} < -60$ km~s\ensuremath{^{-1}}
\item Reject fits with:
\begin{itemize}
\item Large $\chi^{2}$ ($\geq 10$)
\item Mean velocity outside the fitting window
\item Very small ($< 10$ km~s\ensuremath{^{-1}}) or large ($>100$ km~s\ensuremath{^{-1}}) FWHM
\item Negative area
\end{itemize}
\end{itemize}

\begin{figure}[tbh]
\begin{center}
\plottwo{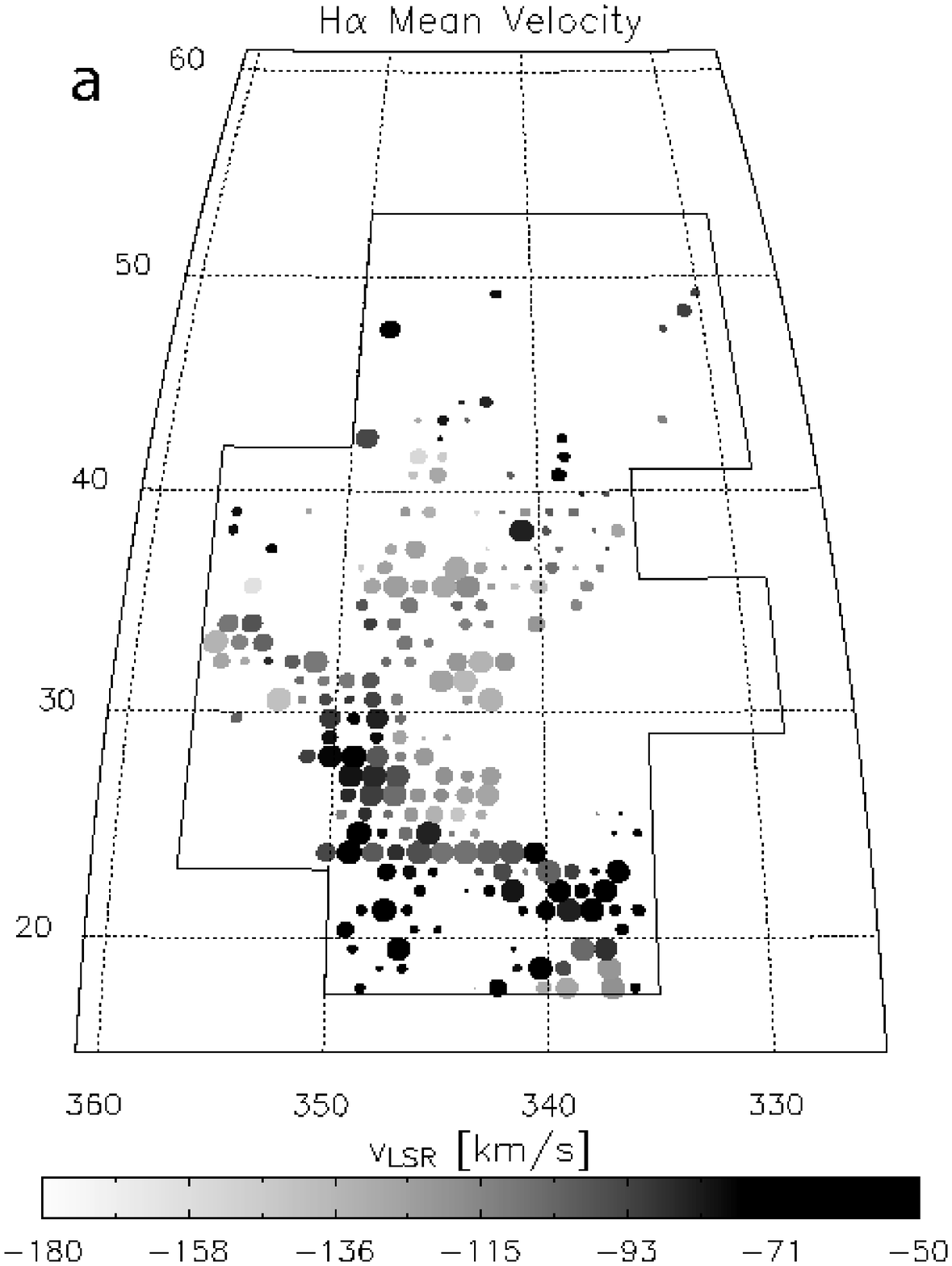}{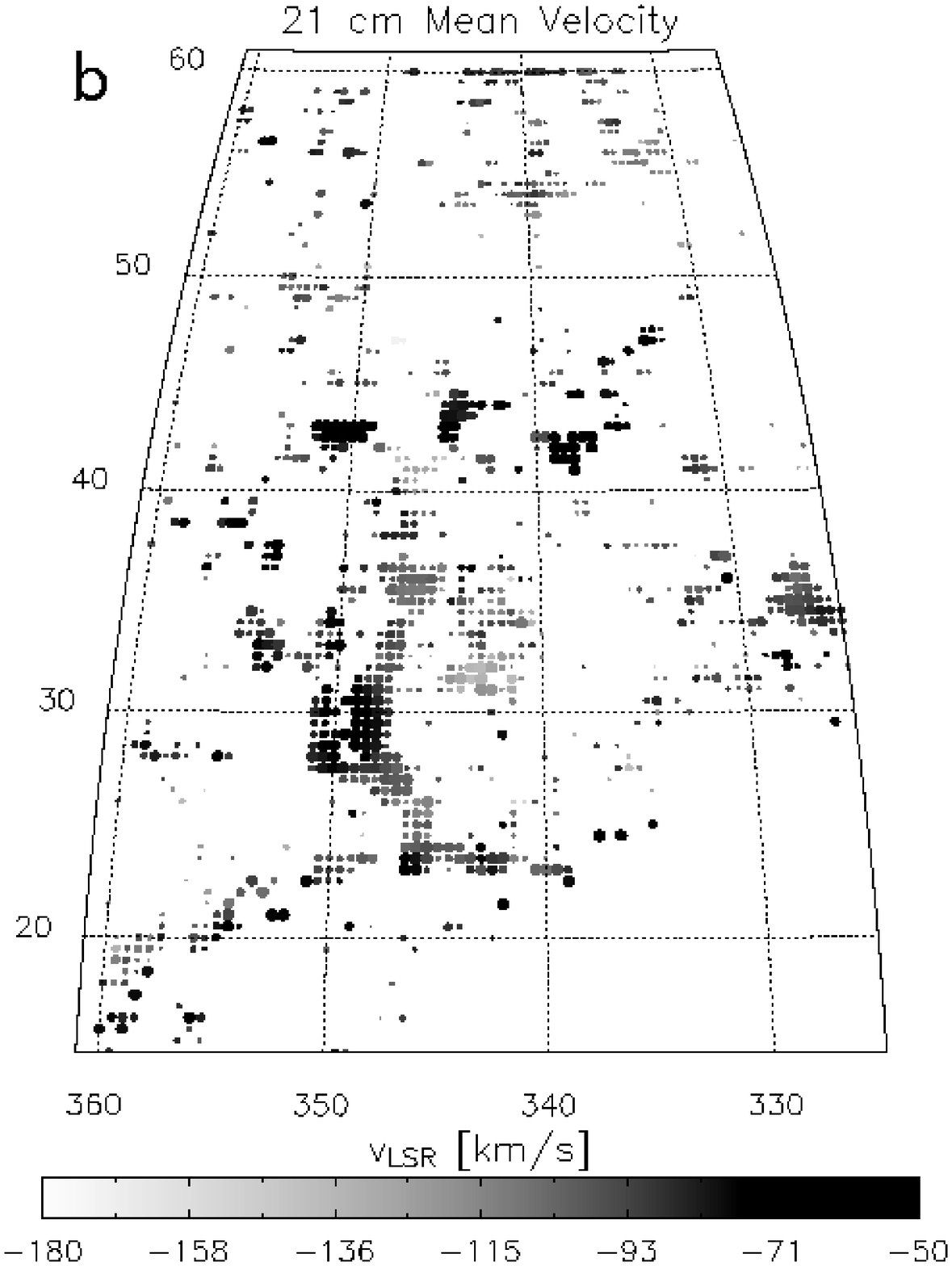}
\caption{H\ensuremath{\alpha}\ and 21 cm mean velocity maps. The grayscale of each beam plotted denotes the central velocity of the emission profile as determined from the Gaussian fit described in the text. The size of each beam is scaled relative to the intensity of the emission with beams containing $\geq 0.3$ R or $\geq 2 \times 10^{19}$ cm$^{-2}$ scaled to 1\deg or 0\fdg5 for \emph{(a)} H\ensuremath{\alpha}\ and \emph{(b)} 21 cm, respectively.}
\label{fig3}
\vspace{-0.5 in}
\end{center}
\end{figure}
 
Using only the newer H\ensuremath{\alpha}\ dataset with the velocity window shifted for better placement of the Complex L emission, about 30\% of the pointings result in ``good'' fits. For the 21 cm dataset, about 15\% satisfy these criteria. Figure~\ref{fig3} shows a representation of how the mean velocities from these fits are distributed across the sky. Here, each emission line is displayed at its original resolution ($1\deg$ for H\ensuremath{\alpha}, $0\fdg5$ for 21 cm) with the size of the plot symbol proportional to the area of the fitted component.

\begin{figure}[tbh]
\begin{center}
\plotfiddle{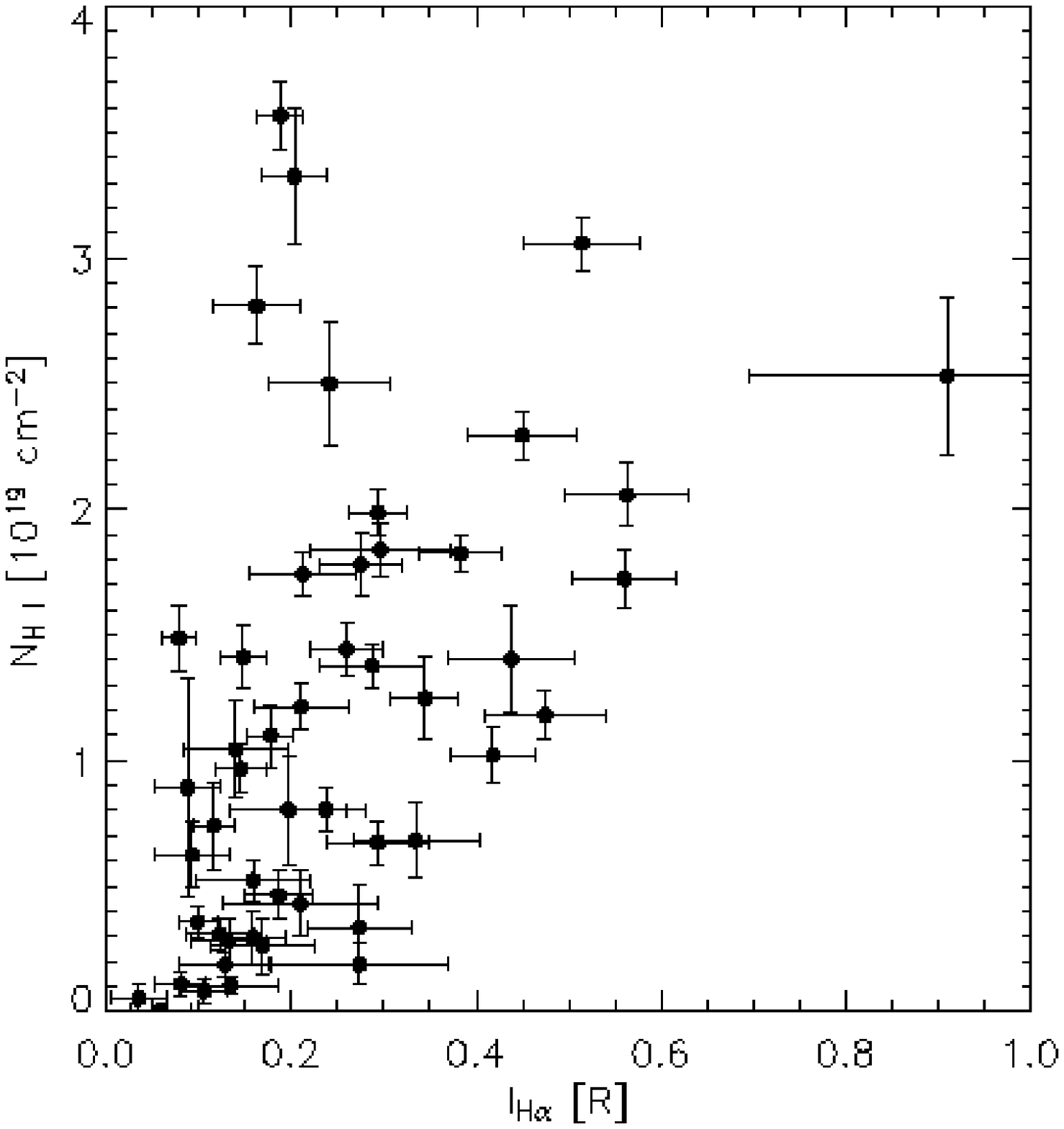}{3 in}{0}{50}{50}{-150}{-90}
\caption{Ionized and neutral component fits: Areas. Column density of \ion{H}{I}\ is plotted versus the intensity of H\ensuremath{\alpha}\ emission for directions in the Complex L region where both are detected. The 21 cm emission has been averaged over the one-degree WHAM beam.}
\label{fig4}
\vspace{-0.35 in}
\end{center}
\end{figure}

\begin{figure}[tbh]
\begin{center}
\plotfiddle{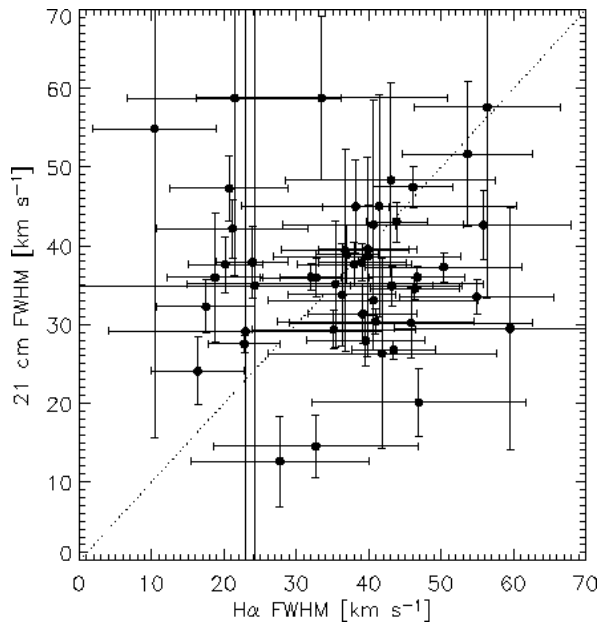}{3 in}{0}{50}{50}{-150}{-90}
\caption{Ionized and neutral component fits: Widths. 21 cm and H\ensuremath{\alpha}\ profile widths are plotted against one another. The substantial error bars are a result of low signal-to-noise profiles from this faint emission. A line of unity is also plotted for reference.}
\label{fig5}
\vspace{-0.3 in}
\end{center}
\end{figure}

\begin{figure}[tbh]
\begin{center}
\plotfiddle{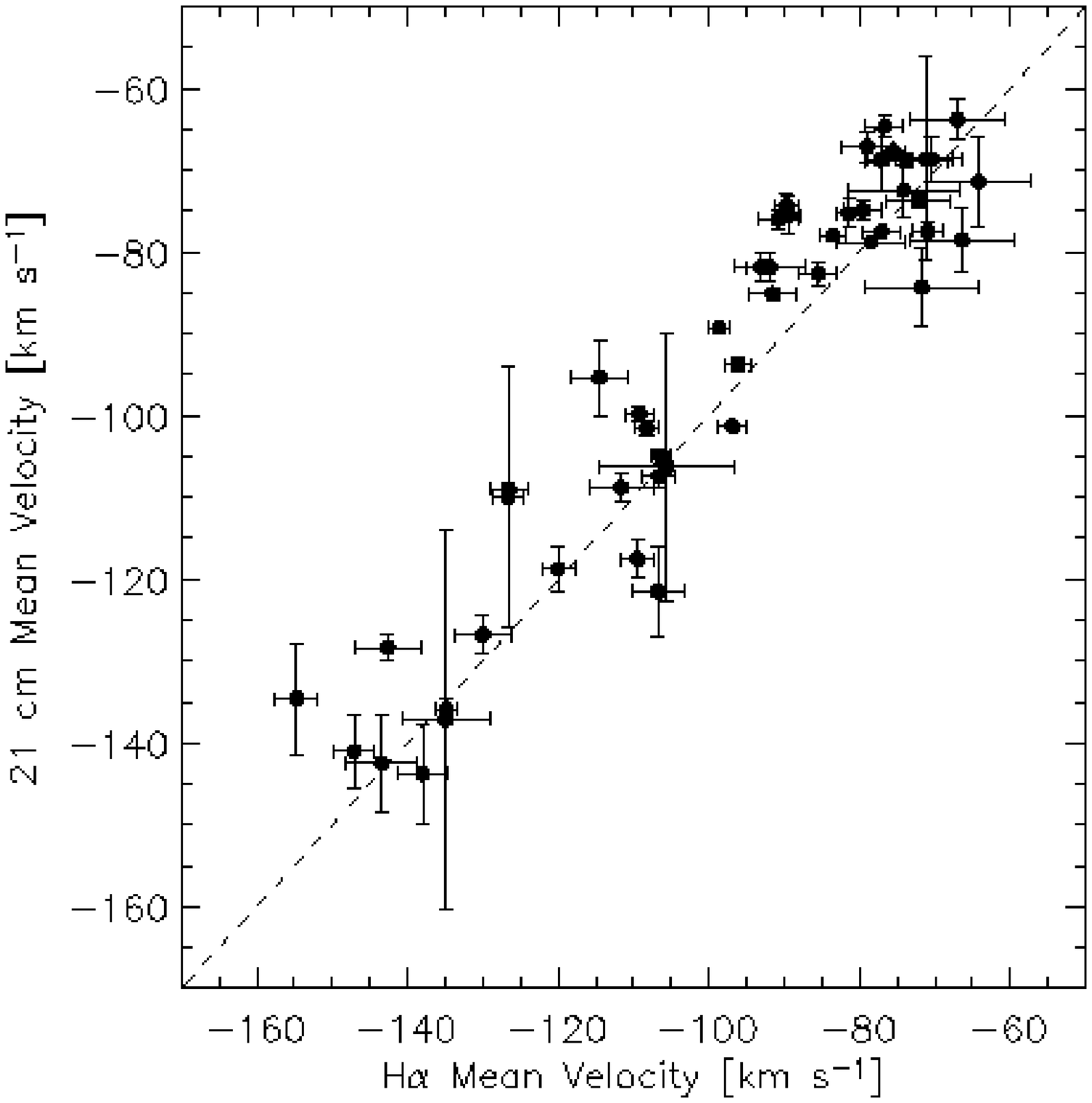}{3 in}{0}{50}{50}{-150}{-90}
\caption{Ionized and neutral component fits: Means. The mean velocity from the Gaussian component fit to the neutral and ionized emission profiles are compared. The dashed line denotes unity.}
\vspace{-0.25 in}
\label{fig6}
\end{center}
\end{figure}

When using 21 cm beams averaged to the match the H\ensuremath{\alpha}\ observations, 49 directions produce ``good'' fits in both emission lines suitable for quantitative comparison. Figures \ref{fig4} through \ref{fig6} plot the 21 cm and H\ensuremath{\alpha}\ values against one another for each of the three Gaussian fit parameters. Figure~\ref{fig4} shows that although there is a general poor correlation between \ensuremath{N_{\mathrm{H\;I}}}\ and \ensuremath{I_{\mathrm{H}\alpha}}, the lack of points in the lower right-hand portion of the diagram and the shape of the lower envelope of the scatter suggest that perhaps there exists some relationship between the two. Empirically, one can interpret this facet of the scatter as a limit on the maximum \ensuremath{I_{\mathrm{H}\alpha}}\ for a given column of neutral gas or that an increase in a minimum \ensuremath{N_{\mathrm{H\;I}}}\ must be present to support the observed increase in H\ensuremath{\alpha}\ brightness. 

Neither statement places strong constraints on a source of ionization for the complex. If the cloud is bathed in radiation by an external source (\emph{e.g.}, the Galaxy or diffuse, cooling halo gas), increased pathlengths along the ionized ``skin'' of the cloud (\emph{i.e.}, lines-of-sight that are not perpendicular to a ionized-neutral boundary) will drive both emission quantities up together. In a scenario where a local source dominates the ionization of the complex, contact between the neutral gas and hot halo material generates a limited amount of ionizing radiation for each such interface. However, lines-of-sight piercing many interface regions can add up to the total neutral and ionized emission observed. The variation in the number of such regions crossed in each observation could provide a base connection between \ensuremath{I_{\mathrm{H}\alpha}}\ and \ensuremath{N_{\mathrm{H\;I}}}\ in those interface regions. However, in either case, the fact that \ensuremath{N_{\mathrm{H\;I}}}\ ranges rather independently from \ensuremath{I_{\mathrm{H}\alpha}}\ beyond this lower envelope suggests that the primary parameters of the ionized region are determined by the details of the ionizing field rather than the physical distribution of the neutral material. 
 
The widths of the components (Figure~\ref{fig5}) trace each other somewhat with a typical FWHM $\sim$ 30--40 km~s\ensuremath{^{-1}}, but the low signal-to-noise of these faint emitting regions prevents much further analysis with this set of observations alone. Such a general value for the width puts an upper limit on the gas temperature of about 35,000 K. Future observations of [\ion{S}{II}] and [\ion{N}{II}] should help constrain this value further.

Finally, Figure~\ref{fig6} shows the close relationship in velocity between the neutral and ionized gas. From our analysis there is a slight shift of the H\ensuremath{\alpha}\ velocity centers compared to the neutral emission. A formal fit gives a shift of $-4.4$ km~s\ensuremath{^{-1}}\ away from the line of unity plotted in Figure~\ref{fig6}. Various checks of the calibration of the H\ensuremath{\alpha}\ velocity scale have been preformed, but a variety of methods show the data have a scale that is correct to better than 1 km~s\ensuremath{^{-1}}. We can confidentially rule this out as an explanation for the shift. The shift between the components also is not a function of any of the Gaussian parameters for either component, suggesting that the fitting routine is not failing systematically in regions of questionable parameter space (\emph{e.g.}, when one of the components becomes very faint). 

Further observations of other ionized species should help to confirm or reject this finding. If this relationship holds, a few scenarios involving ionizing processes do allow for small kinematic shifts of the neutral and ionized components. The velocity suggested here is too low to be indicative of a shock, however conductive interfaces and turbulent mixing layers could result in small relative motions \citep*{SSB93}. On the other hand, external photoionization can also produce a dynamical effect as the ionized gas flows off the face of the neutral cloud \citep[\emph{i.e.}, the ``rocket-effect'';][]{OS55}. In the case of escaping ionizing radiation from the Galaxy, the geometry of IVCs and HVCs targets may be quite conducive to revealing such a process as a blue-shifted signature since the observer and ionizing source lie in approximately the same direction relative to the clouds.
 
\section{Summary}

We have presented the first map of the ionized gas in the intermediate-velocity cloud Complex L. The ionized gas is very clearly spatially and kinematically linked to the neutral gas. However, the intensity of the H\ensuremath{\alpha}\ emission arising from the ionized component of the complex varies independently from the 21 cm emission of the neutral component. Interesting trends including a small systematic offset in the mean velocity of each component and a hints of a relationship in a large sample of \ensuremath{I_{\mathrm{H}\alpha}}\ and \ensuremath{N_{\mathrm{H\;I}}}\ measurements may start to provide clues about the details of what sustains the ionization in IVCs and HVCs. 

In addition to mapping out other complexes in H\ensuremath{\alpha}, we will soon begin exploring the physical conditions in the ionized component of these halo clouds by leveraging the multi-wavelength capabilities of WHAM.

\acknowledgements 

This work is supported by the National Science Foundation through grant AST 02-04973, with assistance from the University of Wisconsin's Graduate School, Department of Astronomy, and Department of Physics.

\end{document}